\documentclass[pra,twocolumn,superscriptaddress]{revtex4}
\usepackage{blindtext}
\usepackage{amsfonts}
\usepackage{amsmath,amssymb}
\usepackage{graphicx}
\usepackage{subfigure}
\usepackage{bm}
\usepackage{epstopdf}
\usepackage{url}
\usepackage{hyperref}
\hypersetup{
     colorlinks = true,
     linkcolor = blue,
     anchorcolor = blue,
     citecolor = blue,
     filecolor = blue,
     urlcolor = blue
}

\begin{document}

\title{Quantum walkers in a disordered lattice with power-law hopping}
\author{G. A. Dom\'{\i}nguez-Castro and R. Paredes} 
\affiliation{Instituto de F\'{\i}sica, Universidad
Nacional Aut\'onoma de M\'exico, Apartado Postal 20-364, M\'exico D.F. 01000, Mexico.}  
\email{rosario@fisica.unam.mx}

\begin{abstract}
We study the effects of interparticle interactions and power-law tunneling couplings on quantum walks executed by both a single one and a pair of hard-core bosons moving in clean and disordered one-dimensional lattices. For this purpose, we perform exact diagonalization to explicitly evaluate the short and long time probabilities of finding the walkers within a surveillance area. Our main conclusions, summarized in phase diagrams in the disorder-power-law and interaction-disorder spaces, allowed us to discern two different scenarios for the single and two quantum walkers dynamics. While in the single particle case the transition to localized and extended regimes is identified for well defined values of the disorder amplitude and power law hopping, those frontiers are replaced by diffuse contours in the interacting two particle case. In fact, counterintuitive transport regimes as diffusion enhanced by disorder, and space constrained dynamics assisted by both interactions and short tunneling range are found. Our results are of direct relevance for  quantum systems with long-range interactions that are currently realized in the laboratory. 
\end{abstract}

\maketitle

\section{Introduction}

Quantum walks, the quantum counterpart of classical random walks \cite{Davidovich1} represent optimal platforms for performing efficient quantum algorithms \cite{Farhi1,Shenvi1}, exploring topological phases \cite{Demler1}, modeling certain photosynthesis processes \cite{Sension1,Lloyd1}, and probing nontrivial dynamics in clean and disordered media in the presence or absence of interactions \cite{Yoav2,Wiater,Chattaraj,Yunbo2,Yunbo1,Yoav1,Paris1}. The wide range of applications that quantum walkers yield, has driven its experimental realization in several platforms such as trapped ions \cite{Roos1,Huber1}, photons in linear and nonlinear waveguides \cite{Crespi1,Kivshar1, White1}, and ultracold atomic gases confined in one-dimensional optical lattices \cite{Karski1, Greiner1, Bloch1}. Due to the high tunability and control that these systems offer, a comprehensive study of a single or many quantum walkers is achievable within the current experimental context.

The dynamics of quantum walkers in disordered media has been widely used for investigating the transport properties of  condensed matter systems \cite{Wiater,Chattaraj, Yunbo2, Yoav1, Roati1,Ghosh, Derevyanko}. As revealed by those studies, the spreading of particles is altered since the disorder breaks the translational symmetry in an otherwise perfectly periodic lattice. For instance, in the one-dimensional Anderson model \cite{Anderson1} any disorder strength yields the exponential localization of the single-particle eigenstates and consequently the absence of particle diffusion. Another widely used model arises in a one-dimensional quasiperiodic system, often called Aubry-Andr\'e (AA) model \cite{Aubry1, Harper, Azbel} (or more precisely, Aubry-Andre-Azbel-Harper model) where the quasiperiodicity emerges as a consequence of superimposing two lattices with incommensurate periods \cite{Dominguez1}. In the AA model, there is a threshold in the disorder strength that signals the transition between extended ergodic and localized single-particle states; therefore, the single-particle diffusion changes from being ballistic, where the spread of the walker grows linearly in time, to becoming null where the particle is constrained to its initial position \cite{Dominguez1}. Both, the Anderson and the AA models are characterized for being tight-binding schemes where the tunneling beyond nearest-neighbors (NN) is exponentially suppressed. This constraint together with the spatial disorder reduces the diffusion of the walkers to the two extreme cases; ballistic and null regimes. To enrich the dynamics exhibited by the walkers, one can replace the nearest-neighbor tunneling with a hopping whose amplitude follows a power-law. This modification is particularly interesting since power-law interactions emerge in many important systems, such as trapped ions \cite{Roos1, Huber1}, polar molecules \cite{Hazzard1, Valtolina1}, Rydberg atoms \cite{Browaeys1}, nuclear spins in solid-state systems \cite{Kaiser1}, photosynthetic complexes \cite{Fleming1, Engel1}, and atoms in photonic crystal waveguides \cite{Cirac1}. Previous analysis of the resulting eigenstates for the disordered quasiperiodic single-particle case have shown that the inclusion of power-law hopping, induces the appearance of energy-dependent mobility edges \cite{Boers1, Santos1} and multifractal states \cite{Santos1}. For the two body case, the literature have focused on the NN tunneling \cite{Orso1, Flach1}, remaining the inclusion of a power law still unexplored.

The purpose of the present manuscript is precisely the study of the interplay among disorder, interparticle interactions, and power law hopping on the spreading of one and two hard-core bosons initially localized in the middle of a one dimensional lattice. The dynamics of a single walker in a disordered lattice with power-law hops is of relevance in quantum state transfer protocols of long-range spin systems \cite{Macri1}, and in quasiparticle propagation experiments on trapped ions \cite{Roos2}. The two-body study is of importance for the understanding of the dynamics of correlated two-particle states \cite{Orso1,Toikka1,Flach1}. In particular, those two-body states whose wave function exhibits multifractality \cite{Diana1}. In addition, two-body transport studies can shed light on more complex phenomena such the many-body localization of interacting long-range systems \cite{Rey1, Santos2, Lev1}.

The main contribution of this investigation is the characterization of different transport regimes exhibited by the quantum walkers.
We performed this study by means of a thorough analysis of the dynamical observables for both, short and long times. We stress that the evolution for long times leads us to predict a transition to localized and extended regimes for the single particle case. In fact, an explicit algebraic dependence of the critical disorder as a function of the hopping power at which the transitions occur was obtained. Regarding the two quantum walkers case, we found a rich dynamics associated with the influence of interactions, disorder and variable tunneling range. For instance, unusual transport regimes such as diffusion stimulated by disorder and space constrained dynamics assisted by both interactions and short tunneling range are detected. 

 
The manuscript is organized as follows. In Sec. \ref{Model} we introduce and briefly discuss the model considered to follow the diffusion of both, the single particle and the two particle cases. Afterwards, in Sec. \ref{Methods} we show the quantities that are calculated during the time evolution of the initial state. The results for the single particle and two interacting particles are shown in Sec. \ref{Results}. Finally, in Sec. \ref{Conclusion} we summarize and conclude the manuscript.


\section{Model}
\label{Model}
We consider quantum walks of both, a single one and two interacting hard-core bosons in a disordered 1D lattice with power-law hopping and periodic boundary conditions. In this scenario, the Hamiltonian that describes the above system is:
\begin{equation}
\begin{split}
H &= -J\sum_{i, j\neq i}\frac{1}{|i-j|^{\alpha}}b^{\dagger}_{i}b_{j} + \Delta\sum_{i}\cos(2\pi\beta i + \phi)n_{i}\\
& + U\sum_{\langle i j \rangle}n_{i}n_{j},
\end{split}
\label{Eq1}
\end{equation} 
where $b_{i}$ ($b^{\dagger}_{i}$) is the bosonic annihilation (creation) operator at site $i$, $n_{i} = b^{\dagger}_{i}b_{i}$ is the corresponding particle number operator, $J$ is the tunneling amplitude between nearest neighbors, and $U$ is the nearest-neighbor interaction energy. Disorder in the lattice is introduced by the second term in Eq. (\ref{Eq1}), in which the parameter $\Delta$ modulates the strength, $\beta=\frac{1+\sqrt{5}}{2}$ is the incommensurable parameter, and $\phi \in [0,2\pi)$ accounts for a random phase. The hard-core constraint implies that no double occupancy is allowed $(b_{i}^{\dagger})^2 = 0$. However, the operators $b_{i}$ and $b_{j}^{\dagger}$ satisfy the usual bosonic commutation relations in different sites $i\neq j$. Here we should point out that the double occupancy restriction emerges naturally in systems where only one excitation per site is allowed. For instance, Rydberg states in neutral atoms, rotational states in polar molecules, and hyperfine states in trapped ions. 

For the noninteracting case $U=0$, and short-range hopping $\alpha\gg 1$, the Hamiltonian in Eq. \eqref{Eq1} approaches to the well known AA model \cite{Aubry1, Harper, Azbel}. Meanwhile, for intermediate $\alpha\gtrsim 1$ and long-range hopping $\alpha<1$, the Hamiltonian in Eq. \eqref{Eq1} is better known as the Generalized-Aubry-Andr\'e (GAA) model \cite{Santos1}. The former, exhibits extended ergodic single-particle states for $\Delta/J < 2$, multifractal at $\Delta/J = 2$, and localized for $\Delta/J>2$. On the contrary, for $\alpha\gtrsim 1$ the GAA model displays a plethora of different regimes with mobility edges, while for long-range hops $\alpha< 1$, it emerges multifractal single-particle states \cite{Santos1}. For the two-body case with nearest-neighbor hopping, it has been shown that within the extended single-particle regime, the interaction enhances the formation of localized pairs \cite{Orso1}. Meanwhile, in the single-particle insulating regime, there is an interaction strength interval in which the two-particle wave function can change from being localized to delocalized \cite{Flach1}. The effects of including power-law tunneling are not yet explored. It is one of the aims of this manuscript to address them. 

\begin{figure}[htbp]
\centering
\includegraphics[width=3.0in]{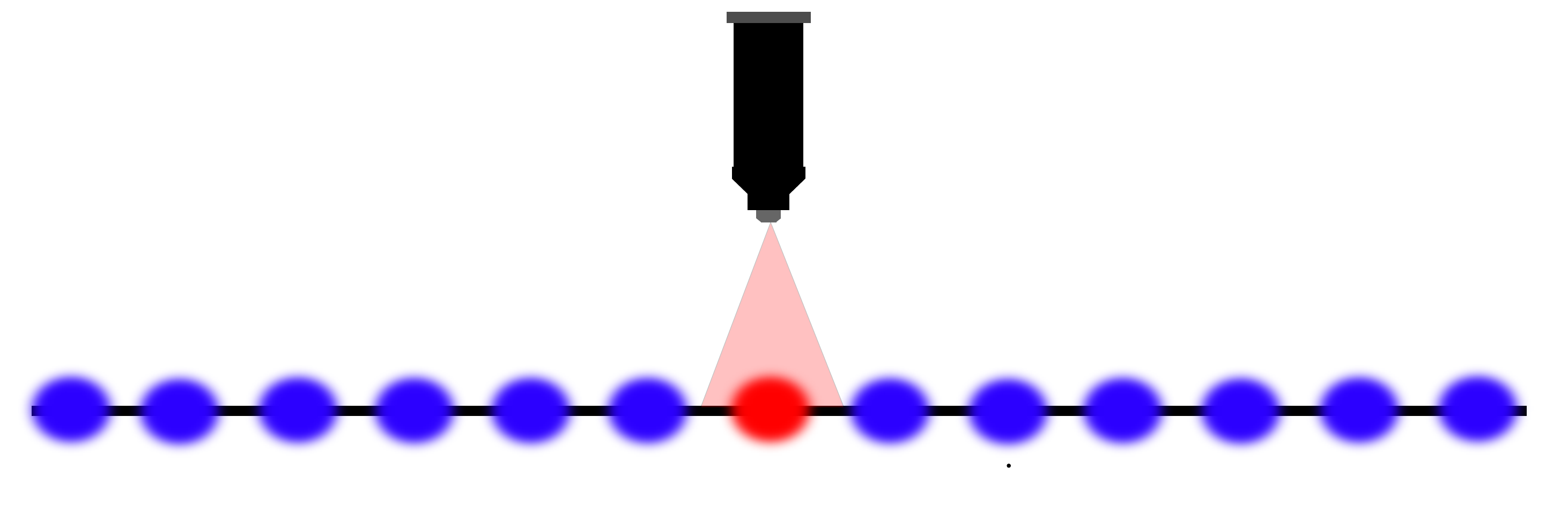}

\vspace{0.2in}

\includegraphics[width=3.0in]{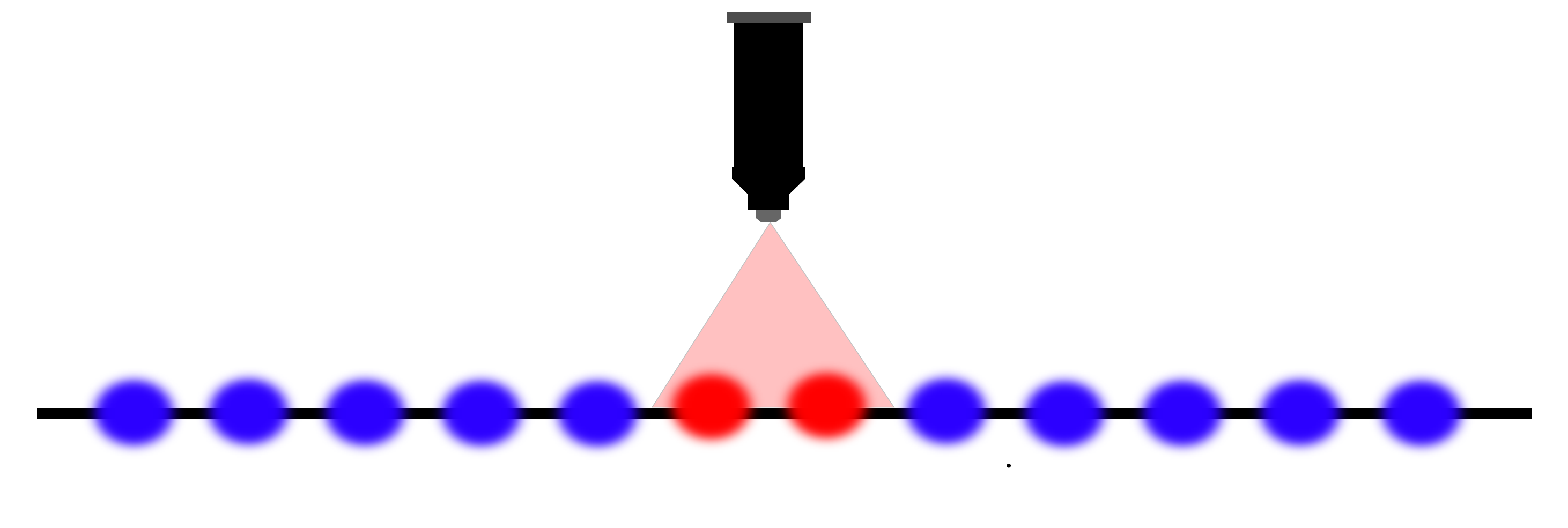}
\caption{Schematic representation of the initial conditions for a single-particle (upper panel) and the two-particle (lower panel) quantum walk.}
\label{Fig1}
\end{figure}

\section{Methods}
\label{Methods}
In Fig. \ref{Fig1} we illustrate the initial states that shall be considered to track the time dynamics of the quantum walkers, that is, one walker at the center of the chain and two walkers localized in adjacent sites in the middle of the lattice. The time evolution of a given initial state $|\psi(t=0)\rangle$ is calculated by using the eigenstates $|\phi_{m}\rangle$ and their corresponding eigenenergies $E_{m}$ obtained from the exact diagonalization of the full Hamiltonian in Eq. (\ref{Eq1})
\begin{equation}
|\psi(t)\rangle = \sum_{m}\langle\phi_{m}|\psi(t=0)\rangle e^{-iE_{m}t/\hbar}|\phi_{m}\rangle.
\label{Eq2}
\end{equation}
During the time evolution of the wave packet $|\psi(t)\rangle$, we monitor the one-particle density $n_{i}(t) = \langle \psi(t)| b_{i}^{\dagger}b_{i}|\psi(t) \rangle$ and the instantaneous survival probability $s(i,t)$ which is defined as the probability of finding a particle within the region $(-i/2,i/2)$ at time $t$,
\begin{equation}
s(i,t) = \sum_{j=-i/2}^{i/2} n_{j}(t).
\label{Eq3}
\end{equation}
In our analysis, we set the surveillance region as half the lattice length $i=L/2$, being $L$ the total number of sites in the lattice. Previous studies for disordered systems \cite{Cuevas1,Torres1} have shown that the instantaneous survival probability provides meaningful information regarding the dynamics of both, noninteracting \cite{Geisel1} and interacting cases \cite{Torres2, Torres3}. In order to condense the entire time evolution of $s(i=L/2, t)$, we calculate its time average
\begin{equation}
\langle s \rangle_{t} = \frac{1}{t}\int_{0}^{t} dt' s(i=L/2,t').
\label{Eq4}
\end{equation}
In the long-time limit, the $\langle s\rangle_{t}$ parameter saturates to the asymptotic survival probability (ASP) $S = \lim_{t \rightarrow \infty} \langle s\rangle_{t}$ \cite{Torres1}, which can be evaluated as follows,
\begin{equation}
S = \sum_{i=-L/4}^{L/4}\sum_{m, \nu}|\langle \phi_{m}|\psi(t=0)\rangle|^{2}|\langle \phi_{m}|n_{1}^{\nu}...n_{\Omega}^{\nu}\rangle|^{2}n_{i}^{\nu},
\label{ASP}
\end{equation}
where $|n_{1}^{\nu}...n_{\Omega}^{\nu}\rangle$ is the occupation number basis, with $\nu$ labeling  the number of elements in the respective Hilbert space. In the single walker case $n_{i}^{\nu}$ takes the simple form $n_{i}^{\nu} = \delta_{i\nu}$ being $\delta_{i\nu}$ the Kronecker delta. The ASP is the main quantity that we shall employ to characterize the transport of the walkers in the long-time limit. In particular, we calculate the ASP for several values of the disorder, interaction and power-law hopping. Due to the fact that measured quantities should not depend on how the disorder is distributed along the lattice, we average each calculation over multiple random realizations of the phase $\phi$ within the interval $[0,2\pi)$.

\section{Results}
\label{Results}

\subsection{Single quantum walker case}
\begin{figure}[h]
\includegraphics[width=0.5\textwidth]{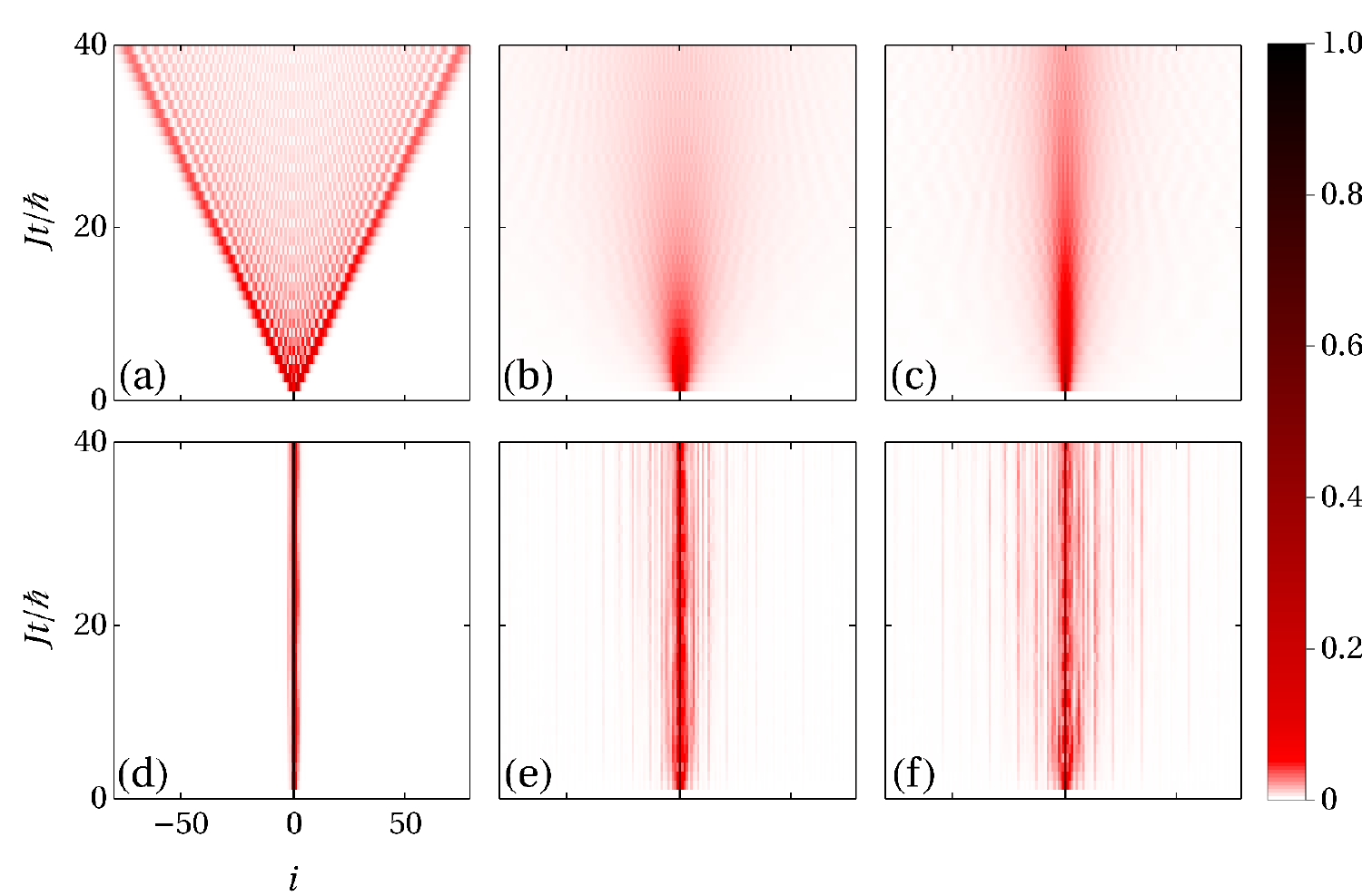}
\caption{Dynamics of the density profile $n_{i}(t)$ of a single quantum walker placed at $t=0$ on the center of a lattice. Hopping powers in panels (a) and (d), (b) and (e), and (c) and (f) are $\alpha=3$, $\alpha=1$ and $\alpha=1/2$ respectively. Upper and lower panels correspond to disorder amplitudes $\Delta/J=0$ and $\Delta/J=8$ respectively.}
\label{Fig2}
\end{figure}

\begin{figure}[h]
\includegraphics[width=0.43\textwidth]{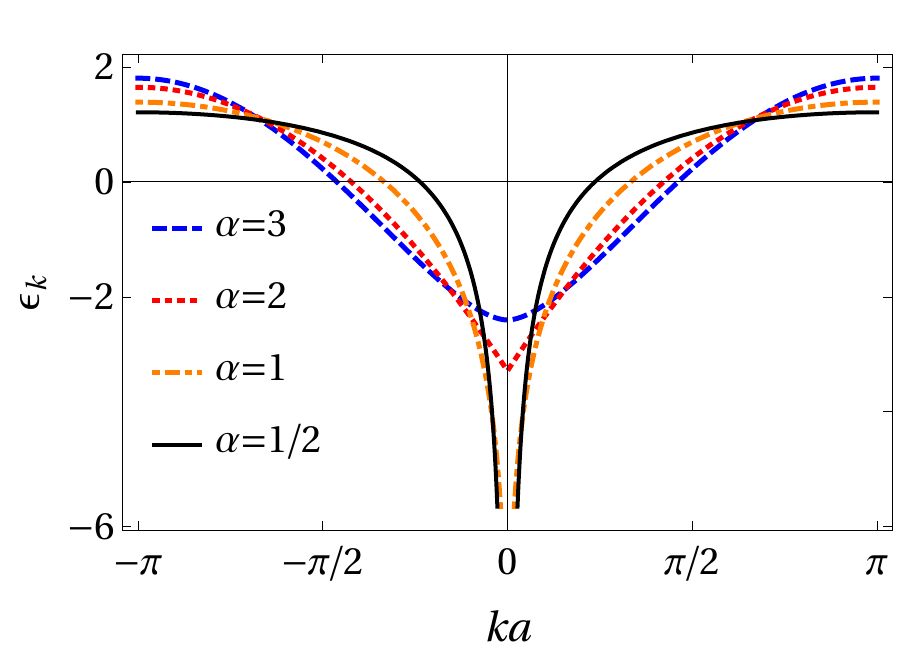}
\caption{Dispersion relation $\epsilon_{k}$ as a function of the lattice momentum $ka$ for hopping powers $\alpha=3$ (dashed-blue), $\alpha=2$ (dotted-red), $\alpha=1$ (dot-dash orange) and $\alpha=1/2$ (solid black).}
\label{Fig3}
\end{figure}

We begin by considering the quantum walk of a single-particle which is initially localized at the center of a lattice with $\Omega=987$ sites. In the upper panels of Fig. \ref{Fig2} we show the time propagation of the one-particle density $n_{i}(t) = \langle b_{i}^{\dagger}b_{i}\rangle$ in the absence of disorder for three different values of the hopping power $\alpha$. For the case $\alpha=3$ (panel (a)) the propagation across the lattice is ballistic, that is, bounded by the group velocity \cite{Yoav1, Kastner1}. In contrast, for the cases $\alpha=1$ and $\alpha=1/2$ (panels (b) and (c)) the notion of group velocity breaks down \cite{Kastner1}, thus originating the peculiar dynamics shown in Fig. \ref{Fig2}, namely, a supersonic-like propagation. To understand this behavior in Fig. \ref{Fig3}, we plot the dispersion relation $\epsilon_{k}$ associated with different values of the hopping power $\alpha$. As one can notice, the derivative of the dispersion relation $\epsilon_{k}$, that is the group velocity, is undefined for $\alpha \leq 2$, and thus resulting in a supersonic-like diffusion. It is worth to mention here that this behavior has been observed in different physical scenarios \cite{Romero-Rochin, Monroe1}. 

For finite disorder strength, remarkable outcomes are also exhibited for the single-particle dynamics. In the lower panels of Fig. \ref{Fig2} we illustrate the average density profile $n_{i}(t)$ of a walk with disorder amplitude $\Delta/J=8$. We averaged over $100$ disorder realizations. Since for the case $\alpha=3$, all single-particle states are localized, the diffusion of the particle is absent. In contrast, for cases $\alpha = 1$ and $\alpha = 1/2$, one can observe that the particle is not as bounded as for the $\alpha = 3$ case. In fact, the one-body density spreads over sites that are not necessarily close to its initial position. This peculiar behavior of covering a sub-extensive number of sites is linked to the existence of multifractal or nonergodic single-particle states \cite{Mirlin}.

\begin{figure}[h]
\includegraphics[width=0.4\textwidth]{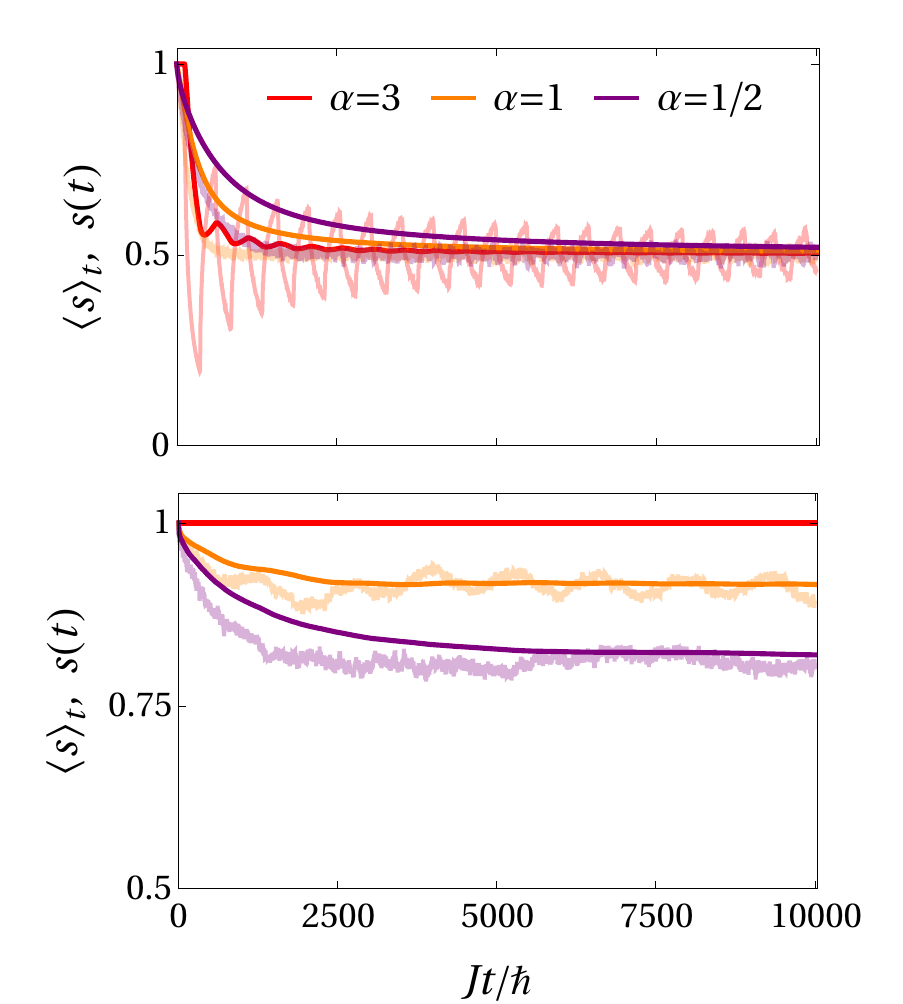}
\caption{Instantaneous survival probability $s(t)$ and its time average $\langle s \rangle_t$ of a single walker placed at $t=0$ on the center of a lattice. Upper and lower panels correspond to disorder amplitudes of $\Delta/J=0$ and $\Delta/J=8$ respectively. Curves with light colors correspond to the instantaneous survival probability $s(t)$, and the darker lines are for the time averages $\langle s\rangle_{t}$. The value of the hopping power $\alpha$ in each curve is indicated within the upper panel.}
\label{FigN}
\end{figure}

To better characterize the spreading patterns shown in Fig. \ref{Fig2}, we calculate the instantaneous survival probability $s(t)$ and its time average until a time of $Jt/\hbar = 10^{4}$. The upper panel in Fig. \ref{FigN} corresponds to zero disorder while the lower panel considers $\Delta/J = 8$. In each panel, the curves in light colors correspond to the instantaneous survival probability $s(t)$, while the darker lines are their corresponding time averages $\langle s\rangle_{t}$. Fig. \ref{FigN} confirms the behavior observed above, that is, for $\Delta/J=0$ the walkers can spread through the whole lattice with, in the average, no preferable site. Therefore, the time average $\langle s\rangle_{t}$ reaches the value $1/2$. On the other side, for $\Delta/J = 8$ and $\alpha = 3$ the walker can not escape from the surveillance region, and for this reason, the probability of finding the particle within this area is $1$. For $\Delta/J = 8$ and $\alpha = 1$ or $\alpha = 1/2$ the walker partially covers sites outside the surveillance area and hence the $\langle s\rangle_{t}$ quantity takes a value between $1/2$ and $1$. Notice that for $\alpha = 1/2$ the walker manages to spread more than for the $\alpha=1$ case.

\begin{figure}[h]
\includegraphics[width=0.5\textwidth]{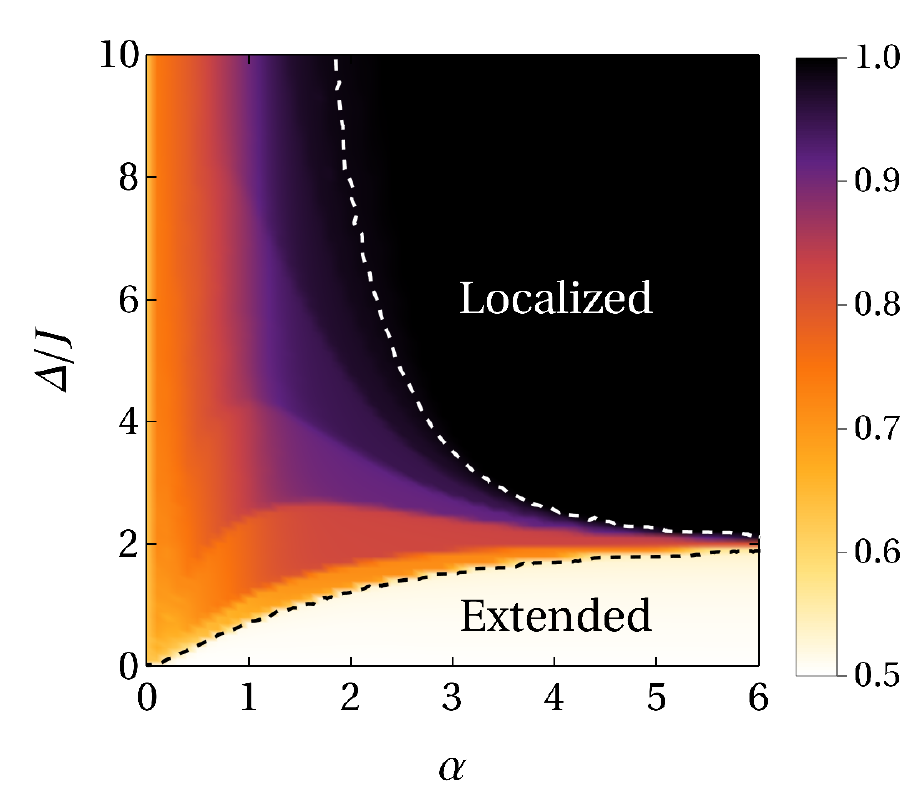}
\caption{Asymptotic value of the survival probability $S$ as a function of the power law $\alpha$ and disorder strength $\Delta/J$. The results are averaged over $100$ disorder realizations.}
\label{Fig5G}
\end{figure}

In Fig. \ref{Fig5G} we summarize the one-body transport exhibited in the GAA model, in particular we show the ASP parameter for different values of the hopping power $\alpha$ and the disorder strength $\Delta/J$.  A noticeable remark from Fig. \ref{Fig5G} is that the survival parameter forecast a smooth transition from the intricate structure of the GAA model to the simple AA model. One can easily recognize the latter by observing how the extended and localized boundaries get closer for $\alpha\gg 1$ and $\Delta/J=2$.
Last but not less important are the different regimes shown within the extended and localized boundaries. For finite $\alpha\gtrsim1$ there are well-defined curves where the particle dynamics shows bimodal behavior, that is, the walker partially escapes and partially remains localized inside the surveillance region. This peculiar transport signals the presence of mobility edges which split the extended and localized eigenstates. As stated by \cite{Santos1}, the amount of extended or localized states depends on both, the power hop $\alpha$ and the disorder strength $\Delta/J$, being more localized states for large disorder strengths. This explains the different numerical values that the $S$ parameter displays. For long-range hops $\alpha<1$, the spectrum of the Hamiltonian in Eq. (\ref{Eq1}) exhibits ergodic and multifractal single-particle states. Consequently, the walker is less constrained to stay inside the observation area than walkers with intermediate hops $\alpha\gtrsim 1$. 

As one can notice, the results in Fig. \ref{Fig5G} are in remarkable agreement with the diagram found in \cite{Santos1}. This suggests that monitoring the asymptotic survival probability of a single quantum walker can signal the inherent structure of the eigenstates of Hamiltonian in Eq. \ref{Eq1}. Furthermore the results in Fig. \ref{Fig5G} provide valuable information with the advantage of being in terms of a mesurable quantity in current experimental setups. 

The critical disorder strength required to localize the system can be well fitted to a power-law function of the hopping range $\alpha$,
\begin{equation}
\frac{\Delta_{c}^{L}(\alpha)}{J} = 2 + a_{1}\alpha^{b_{1}}.
\label{Eqnew1}
\end{equation}
From Fig. \ref{Fig5G}, one can anticipate that the value of $b_{1}$ must be negative since the critical disorder has to approach the well-known $\Delta_{c}^{L}/J = 2$ transition value for short range hops. Our calculations reveal that the localization transition for the GAA follows Eq. \ref{Eqnew1} with $a_{1} = 64.3 \pm 3.5$ and $b_{1} = -3.43 \pm 0.04$. Analogously, we can find an expression for the disorder strength $\Delta_{c}^{E}$ for the extended transition. In this case, the disorder magnitude appears to follow the behavior of a rational function
\begin{equation}
\frac{\Delta_{c}^{E}(\alpha)}{2J} = \frac{\alpha^{b_{2}}}{a_{2}+\alpha^{b_{2}}}.
\label{Eqnew2}
\end{equation}
We obtain that $a_{2} = 2.0 \pm 0.02$ and $b_{2} = 1.63 \pm 0.03$. Summarizing, the expressions in Eqs. \ref{Eqnew1} and \ref{Eqnew2} yield the value of the disorder at which the walker shows an absence of diffusion or it is allowed to spread over the whole lattice. Thus, generalizing the results for NN hopping since we have considered short and long range tunneling couplings. The physical origin of $a$'s and $b$'s parameters is beyond the scope of this work and will be the subject of future studies  

\subsection{Two interacting quantum walkers case} 

In order to analyze the interplay among interactions, disorder, and power-law hopping, we now consider the quantum walk of a pair of interacting hard-core bosons. As stated in section \ref{Methods}, the particles are initially placed in the middle of a chain having $\Omega = 62$ sites. All the results in this subsection correspond to the average over 400 realizations of the random phase $\phi$. 

\begin{figure}[h]
\includegraphics[width=0.5\textwidth]{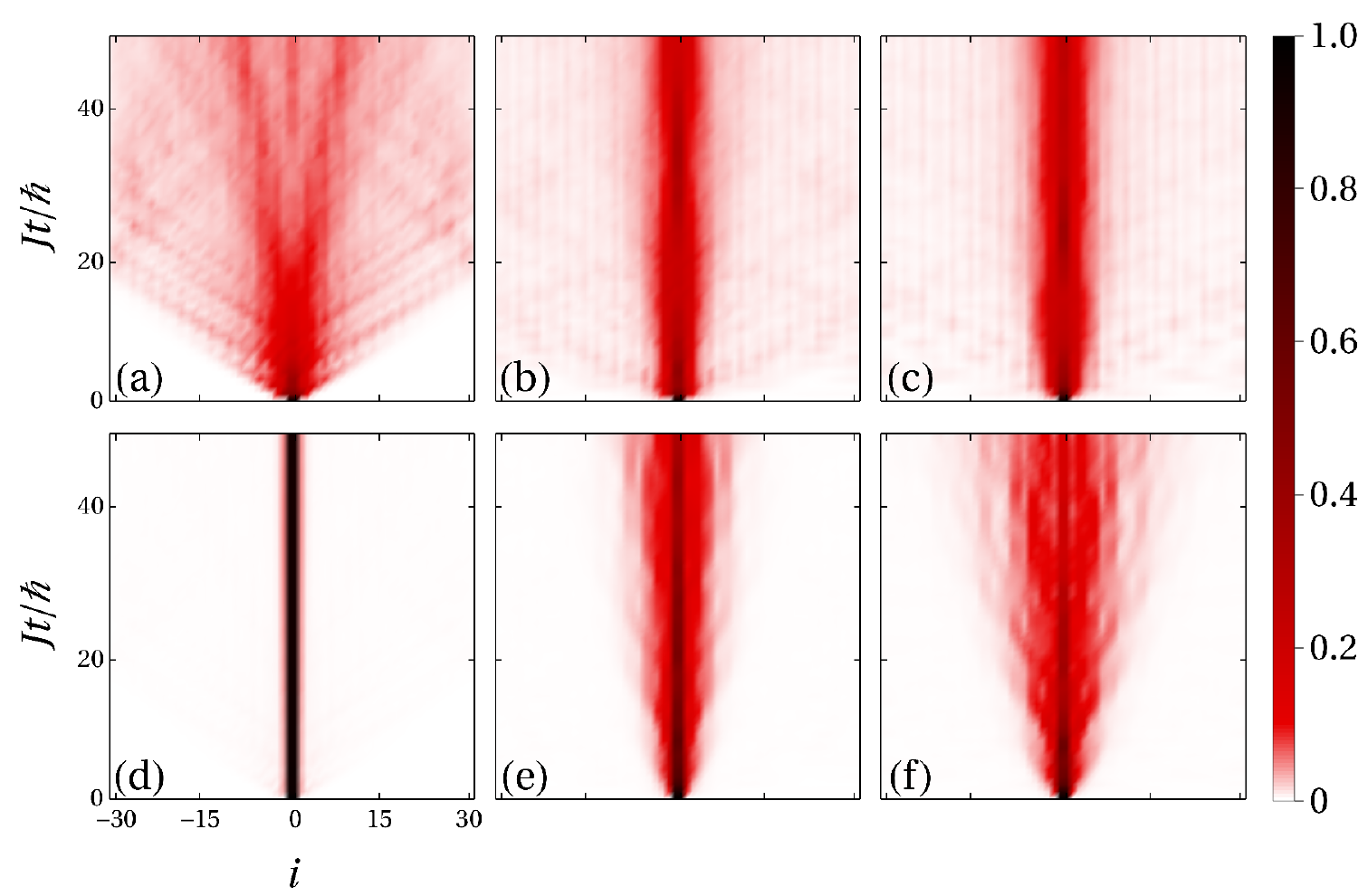}
\caption{Dynamics of the density distribution $n_{i}(t)$ of two quantum walkers placed at $t=0$ on the center of a disordered lattice, with disorder amplitude $\Delta/J=1$, hopping powers $\alpha=3$ (left) $\alpha=1$ (center) and $\alpha=1/2$ (right), and fixed nearest-neighbor interactions $U/J=1$ (upper panels) and $U/J=8$ (lower panels).}
\label{Fig6G}
\end{figure}

\begin{figure}[h]
\includegraphics[width=0.5\textwidth]{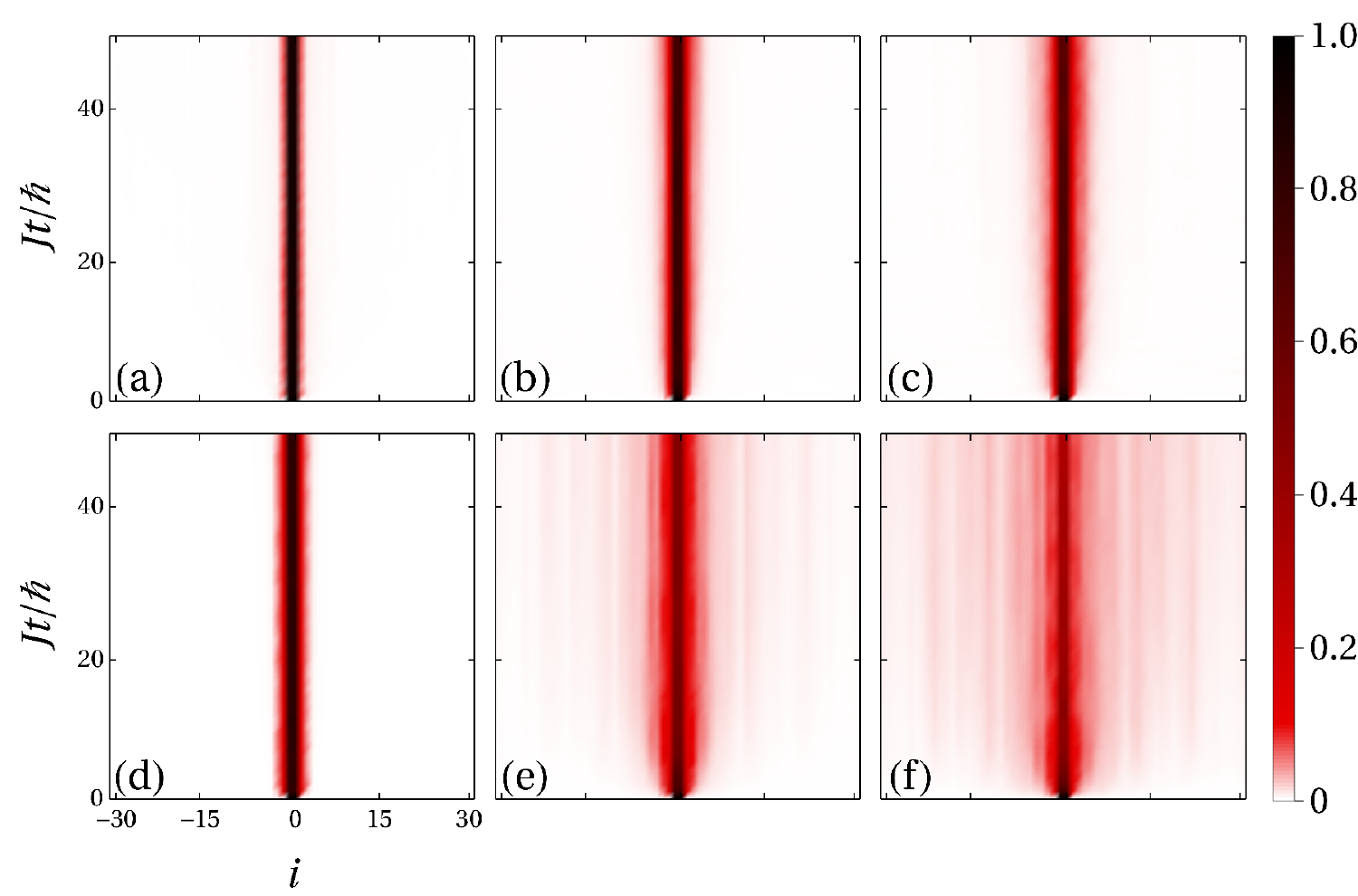}
\caption{Dynamics of the density distribution $n_{i}(t)$ of two quantum walkers placed at $t=0$  on the center of a disordered lattice, with fixed interaction strength $U/J=8$, hopping powers $\alpha=3$ (left) $\alpha=1$ (center) and $\alpha=1/2$ (right), and fixed nearest-neighbor interactions $\Delta/J=2$ (upper panels) and $\Delta/J=8$ (lower panels).}
\label{Fig7G}
\end{figure}

\begin{figure}[h]
\includegraphics[width=0.5\textwidth]{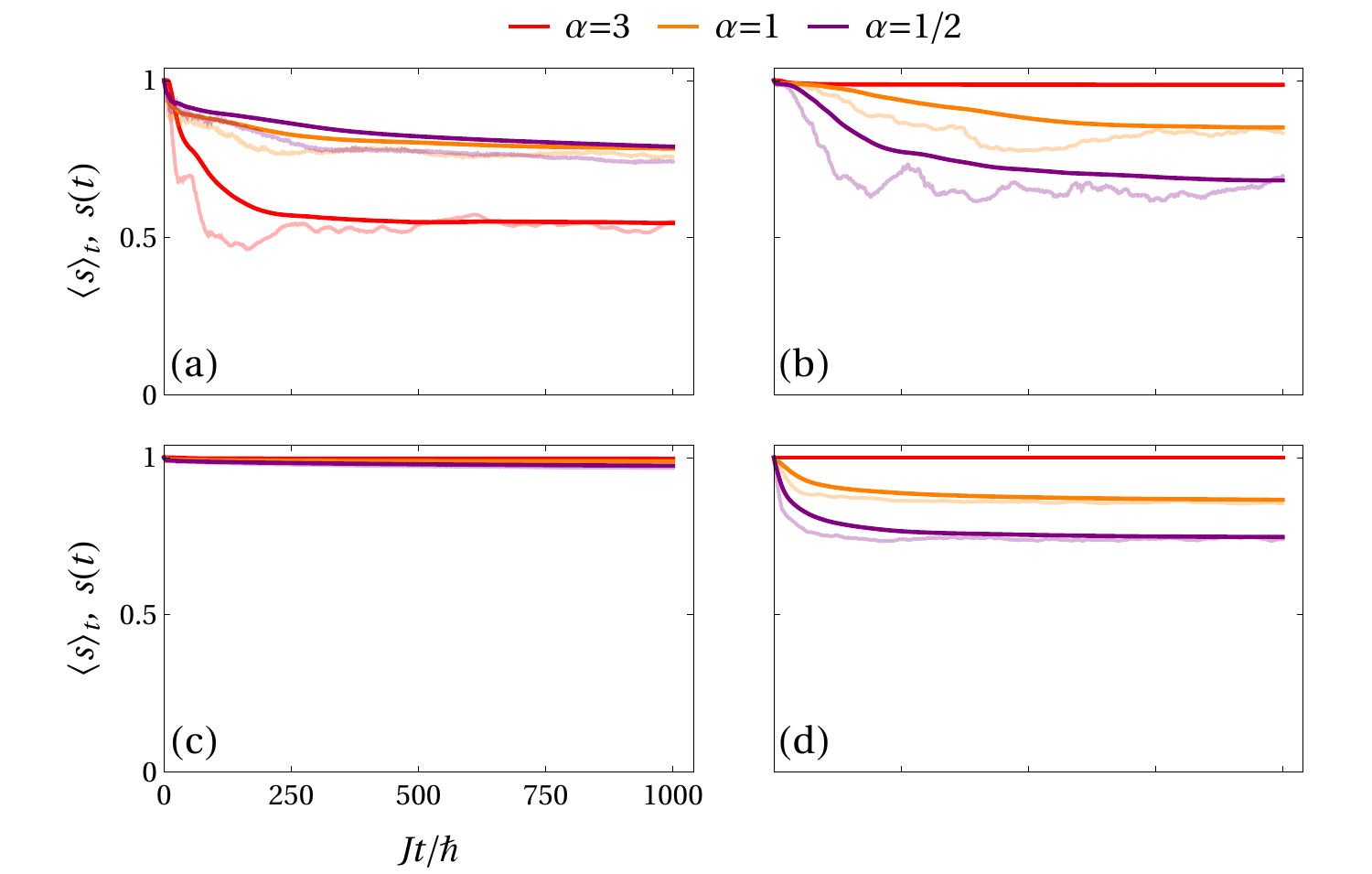}
\caption{Instantaneous survival probability $s(t)$ and its time average $\langle s \rangle_t$ of two walkers placed at $t=0$ on the center of a lattice. Curves with light colors correspond to the disorder average of the instantaneous survival probability $s(t)$, and darker lines are for the time averages $\langle s\rangle_{t}$. The value of the hopping power $\alpha$ in each curve is indicated in the top of the figures. Left and right panels correspond to values of disorder and interaction strengths  indicated in figures \ref{Fig6G} and \ref{Fig7G} respectively.}
\label{Fig8G}
\end{figure}

In Fig. \ref{Fig6G} we show the time evolution of the single-particle density $n_i$ for three different values of the hopping power $\alpha$ (left, center and right panels correspond to $\alpha=3,1$ and 1/2 respectively), and a fixed disorder strength of $\Delta/J = 1$. Upper panels ((a), (b) and (c)) correspond to $U/J = 1$ while lower panels ((d), (e) and (f)) consider $U/J = 8$. 
The dynamics of the walkers for $U/J=1$ is reminiscent to the single-particle case since the pair of particles with $\alpha=3$ manage to spread over more sites outside the surveillance area than the walkers with $\alpha=1$ and $\alpha=1/2$. However, in the lower panels of Fig. \ref{Fig6G} one can appreciate how this previous scenario is reversed as the interaction strength increases. Thus revealing that the diffusion of walkers with long-range hops is robust to strong interactions.
Fig. \ref{Fig7G} is the analogous of \ref{Fig6G}, that is, the behavior of the time-dependent density profile $n_{i}(t)$ for three different values of the hopping power $\alpha$, but having a fixed interaction strength of $U/J = 8$. The upper panels ((a), (b) and (c)) consider $\Delta/J = 2$ while the lower ones ((d), (e) and (f)) correspond to $\Delta/J = 8$. A remarkable feature emerging from Fig. \ref{Fig7G} is the fact that for $\alpha=1$ and $\alpha=1/2$  the density profile achieves to expand to greater distances as the disorder magnitude is increased. Thus, contradicting the general notion that the disorder yields suppression of transport.
To quantify the spreading patterns shown in Figs. \ref{Fig6G} and \ref{Fig7G}, we evaluate the instantaneous survival probability $s(t)$ and its time average $\langle s\rangle_{t}$ until a time of $Jt/\hbar = 10^{3}$. The upper panels in Fig. \ref{Fig8G} correspond to the cases shown in Fig. \ref{Fig6G} while the lower panels are associated with the cases displayed in Fig. \ref{Fig7G}. One can confirm how the interplay between interactions and disorder gives rise to an unexpected behavior, namely, that a larger disorder amplitude produces greater spreading than those accounted for lower values of disorder.

\begin{figure}[h]
\includegraphics[width=0.5\textwidth]{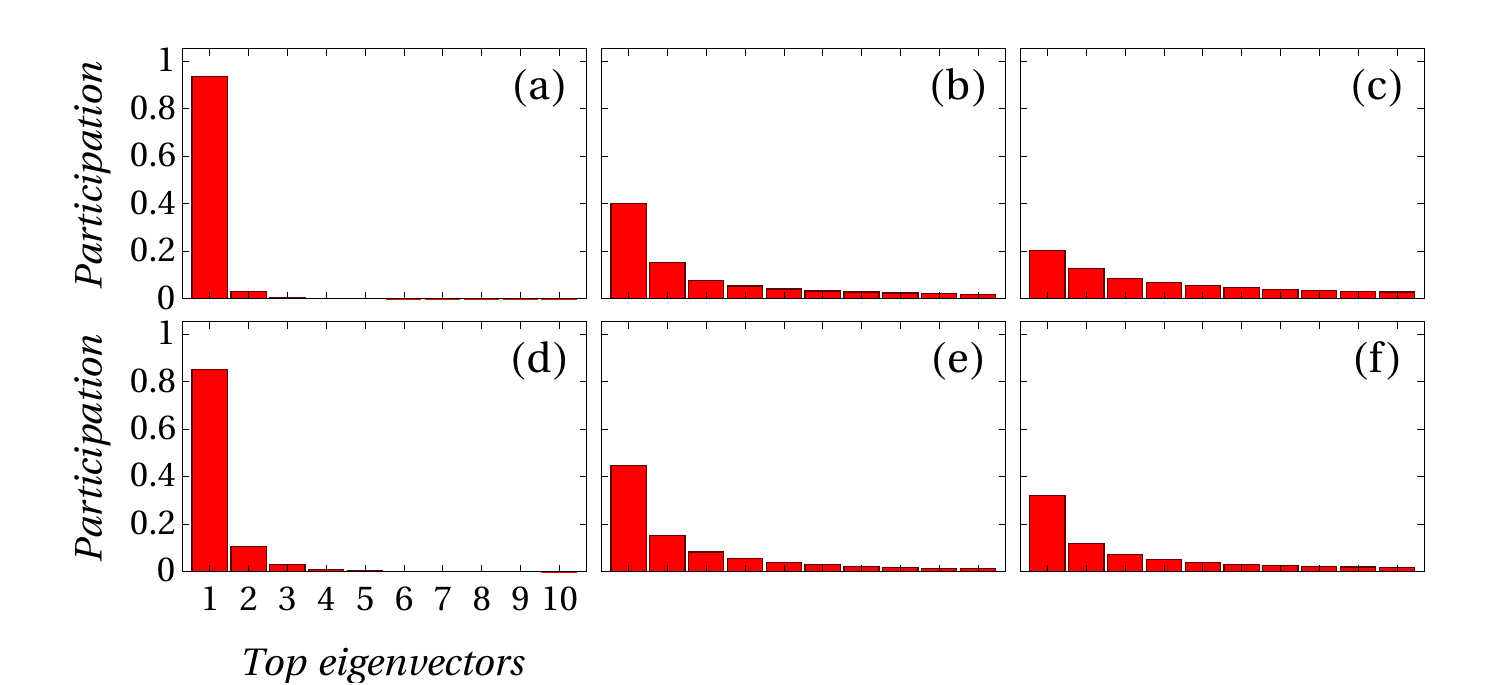}
\caption{Ten eigenstates with larger contribution to the coefficients in equation \ref{Eq2} for three hopping powers $\alpha=3$ (left) $\alpha=1$ (center) and $\alpha=1/2$ (right). Upper panels correspond to interaction strength $U/J=8$ and disorder amplitude $\Delta/J=1$. Lower panels are associated with interaction strength $U/J=8$ and disorder amplitude $\Delta/J=8$. The results are averaged over $400$ disorder realizations.}
\label{Fig9G}
\end{figure}

A way to understand the behavior of the density dynamics exhibited in Figs. \ref{Fig6G} and \ref{Fig7G}, is by means of the analysis of the eigenstates contribution to the linear combination that builds the initial state. In Fig. \ref{Fig9G}, we show the extent of the coefficient $|\langle \phi_{m}|\psi(t=0)\rangle|^2$ associated with the ten eigenvectors that contribute the most to the superposition in Eq. \ref{Eq2}. All the panels in this figure correspond to $U/J=8$, while the upper ones are associated with $\Delta/J=1$ the lower panels correspond to  $\Delta/J=8$. 
As can be appreciated from Fig. \ref{Fig9G}, for $\alpha=3$ we observe an almost single eigenstate dominated contribution, while a collective eigenstate participation for $\alpha=1$ and $\alpha = 1/2$ dictates the lattice dynamics. It is worth to mention that the $x$-axis label in Fig. \ref{Fig9G} indicates the hierarchical order of the contribution and not the eigenvector index.

\begin{figure}[h]
\includegraphics[width=0.48\textwidth]{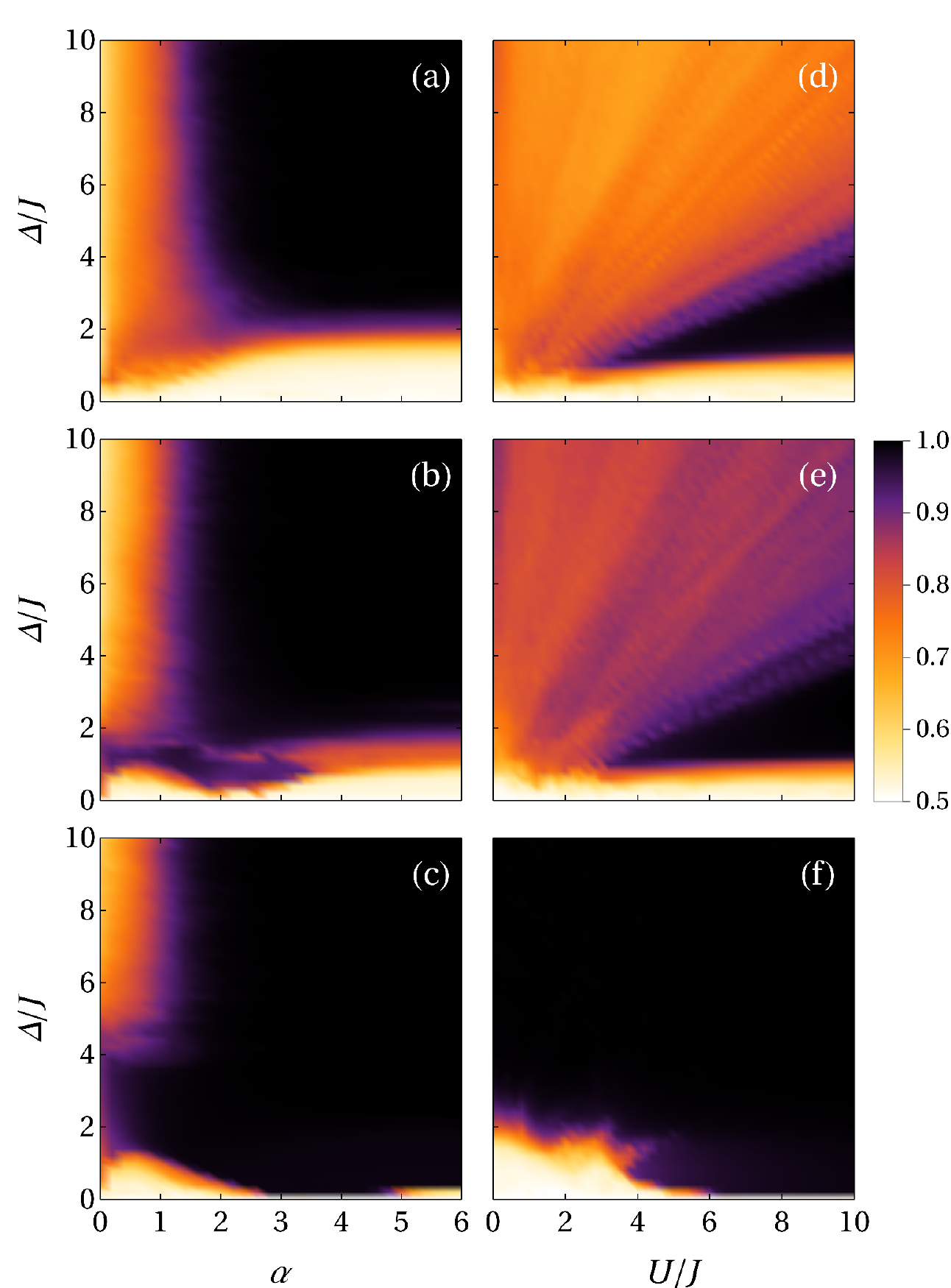}
\caption{Asymptotic value of the survival probability $F(i=\Omega/2)$ in the long-time limit. Left panels correspond to a parameter space of disorder strength vs. hopping power, $\Delta/J$ vs. $\alpha$, and right panels illustrate $S$ as a function of the disorder amplitude against nearest neighbor interaction, $\Delta/J$ vs. $U/J$. The results are averaged over $400$ disorder realizations.}
\label{Fig10}
\end{figure}

Analogously to the study for the single quantum walker, we track the behavior of the long time dynamics through the asymptotic survival probability $S$ for two quantum walkers. In Fig. \ref{Fig10} we condense in a density color scheme the full information associated with both, the competition between disorder and hopping power, and the interplay between disorder and interactions. Similarly to Fig. \ref{Fig5G} black and white colors are associated with the opposite localized and extended regimes respectively. We point out that for convenience, we rescaled the two-body ASP parameter by a factor of 1/2 to match the scale in Fig. \ref{Fig5G}. From top to bottom, left panels illustrate the behavior of $S$ associated with $U/J =1,4,$ and $8$ respectively (see Figs. \ref{Fig10}(a), \ref{Fig10}(b) and \ref{Fig10}(c)), while the right panels correspond to power hops $\alpha=1/2, 1,$ and $3$ respectively (see Figs. \ref{Fig10}(a), \ref{Fig10}(b) and \ref{Fig10}(c)). As can be seen from Fig. \ref{Fig10}(a), the dynamics for the weakly interacting case, $U/J=1$, resembles the behavior found for the single walker case. Both, the smooth frontiers for $\alpha <3$ and the well defined contours identifying the transition to localized and extended regimes shown in Fig. \ref{Fig5G}, vanish when interaction is turned on. However, it still is possible to discern clearly the extended and localized regions in the space of parameters. As the interaction amplitude is increased, $U/J=4$, the previous diagram with undefined frontiers turns into blurred zones in which the extended regimes begin to disappear for values of the disorder strength smaller than $\Delta/J<2$.
It is important to stress that for the largest interaction considered $U/J = 8$, the pair of particles experience no restricted motion for small values of disorder and long-range hopping. Remarkably, the intermediate transport region in which the walkers escape and remain localized partially within the surveillance area,
 endures strong interactions, where it gets shifted to larger values of the disorder strength. Regarding the interplay between disorder and interactions $\Delta/J$ vs $U/J$, one can see from panels on the right of Fig. \ref{Fig10}, that different scenarios are displayed. We notice from panel (f), that for short-range hops ($\alpha=3$), the spread of the particles is clearly jeopardized by disorder and interaction. Consequently, the walkers can only escape the surveillance area for a small region of the $\Delta/J$ vs $U/J$ space. In contrast, panels (d) and (e) exhibit a richer structure where the disorder can either, suppress or stimulate the diffusion of the walkers. 
As one can see from the right panels of Fig. \ref{Fig10}, for short tunneling hops (panel (f)), intermediate values of the interaction strength lead to localized regimes, whereas increasing the tunneling range (see Figs. (d) and (e)) produces extended regimes for arbitrarily large interaction amplitudes, where also the distinctive $\Delta/J=2$ does not prevail any more.

\section{Conclusions}
\label{Conclusion}

We investigated the dynamics of both, a single- and two- hard-core bosons bosons moving in clean and disordered one-dimensional lattices with power-law tunneling $1/r^{\alpha}$. By means of exact diagonalization we tracked the density dynamics and the probability of finding the particles within a surveillance area. Through the study of these observables for short and long times we addressed the effects of the three factors, tunneling range, interactions, and disorder, on the dynamics of the quantum walkers. Besides anticipating the origin of future regimes, the dynamics for short times allowed us to discern ballistic and supersonic-like propagations (for $\alpha\geq 3$ and $\alpha\leq 3$ in the single particle case respectively), as well as finding out unexpected results in both, single and two particle cases; namely, diffusion enhanced by disorder and localized density associated to short disorder but large interaction amplitudes. The whole information associated to the asymptotic behavior of the density spreading across the lattice was summarized in diagrams for the disorder-power-law and interaction-disorder spaces. In the single particle case the transition to localized and extended regimes was captured in the disorder vs. power law space as a well define region. In contrast, we found for the two-particle dynamics that when interactions are turned on, the well defined contours identifying the transition to localized and extended regimes transform into blurred zones in which the strong influence of the power law tunneling give rise to extended transport regimes. 

We expect that our work will trigger further theoretical analysis such as determining the fractal nature of the two-body states in intermediate and long-range hops by means of projected Greens function methods \cite{Diana1}. Another striking problem is the fate of the disorder-induced spreading within the many-body regime. Our results are of interest for experiments with trapped ions, Rydberg atoms, and photons in crystal waveguides where exotic transport phenomena with long-range interactions are explored.

\acknowledgments{GADC would like to thank Gerardo G. Naumis for useful discussion and V. Romero-Roch\'in for computational resources at Skyrmion IF-UNAM. This work was partially funded by grant  IN108620 DGAPA (UNAM). GADC acknowledges CONACYT scholarship.
}

\end{document}